\RequirePackage{fix-cm}
\documentclass[twocolumn]{svjour3}

\usepackage{graphicx}
\usepackage{amsmath}

\usepackage{hyperref}
\usepackage{booktabs}
\usepackage{tabularx}
\usepackage{subfigure}
\journalname{Granular Matter}
\usepackage{xcolor}

\renewcommand{\vec}[1]{\boldsymbol{#1}}

\begin{document}
\title{DEM-Simulation of thin elastic membranes interacting with a granulate}
\author{Holger Götz \and Thorsten Pöschel}
\institute{Holger G\"otz \and Thorsten P\"oschel\at
Institute for Multiscale Simulations\\
Friedrich-Alexander-Universit\"at Erlangen-N\"urnberg\\
Cauerstra\ss{}e 3, 91058 Erlangen\\
Germany\\
\email{holger.goetz@fau.de}           
}
\date{Received: \today / Accepted: date}
\maketitle

\begin{abstract}
For a wide range of applications, we need DEM simulations of granular matter in contact with elastic flexible boundaries. We present a novel method to describe the interaction between granular particles and a flexible elastic membrane. Here, the standard mass-spring model approach is supplemented by surface patches given by a triangulation of the membrane. In contrast to standard mass-spring models, our simulation method allows for an efficient simulation even for large particle size dispersion. The novel method allows coarsening of the mass-spring system leading to a substantial increase of computation efficiency. The simulation method is demonstrated and benchmarked for a triaxial test. 
    \keywords{DEM \and boundary condition \and elastic membrane \and triaxial test}
\end{abstract}

\section{Introduction}

For many applications in granular matter research, the system boundaries are given by deformable containers which may be modeled as elastic membranes. Of particular interest are jamming systems where the granulate changes its mechanical properties drastically when the particle number density in the system is changed \cite{liuJammingNotJust1998,ohernJammingZeroTemperature2003,ciamarraRecentResultsJamming2010}, which is frequently achieved by evacuating the air from a deformable container partly filled by granulate. Prominent examples are granular robotic grippers \cite{brownUniversalRoboticGripper2010},
where this mechanism is used to grip and manipulate objects, granular paws \cite{hauserFrictionDampingCompliant2016} and similar \cite{fitzgeraldReviewJammingActuation2020}.

In these cases, the system dynamics is determined by two-way coupling, that is, the deformation of the membrane (e.g. caused by external air pressure) implies forces on the granular particles and the membrane is deformed under the action of the granular packing.
Such membranes can be modeled as \textit{mass-spring systems} (MSS), e.g. \cite{debonoDiscreteElementModelling2012}. Here the membrane is described as a regular or irregular graph whose vertices are particles and whose edges are linear or non-linear elastic springs. The elastic behavior is, thus, described by the springs and the contacts between the membrane and the enclosed granular particles are described by the contacts between the membrane's particles and the particles of the granulate.

The choice of the membrane's structure is critical: If the meshes are too coarse, particles can penetrate the membrane. Therefore, the mesh width has to be chosen according to the smallest particles in the system \cite{quDiscreteElementModelling2019,debonoDiscreteElementModelling2012}. This is problematic for several reasons: First, in the coarse of the simulation when the membrane and the granulate particles interact, the mesh width may change which is difficult to predict. Second, for the simulation of a highly disperse granulate, the number of membrane particles and springs can be very large, resulting in an inefficient simulation. The problem can be solved by artificially increasing the sizes of the non-interacting membrane particles (overlapping particles) \cite{wuStudyShearBehavior2021} which, however, introduces an undesired thickness of the membrane. All MSS models are problematic for the modeling of tangential (frictional) forces along the membrane since the contacts of the membrane particles - granular particles depends on the concrete arrangement of the particle positions.

In the current paper, we describe a novel type of MSS which allows for the simulation of impenetrable flexible elastic boundaries requiring a moderate number of membrane particles. This model does not suffer from the drawbacks discussed above. Since by construction the membrane is impenetrable, the meshes can be chosen larger which makes our method computationally efficient. The proposed model was used recently to simulate a granular gripper \cite{gotzSoftParticlesReinforce2021} and a bending beam of granular meta-material \cite{bendingBeamMeta:2022}.

\section{Model description}
\label{sec:model-description}

\subsection{Particle: particle interaction}
\label{sec:part-part-inter}

The discrete element method (DEM) solves Newton's equation for the position $\vec{r}_i$ and the angular orientation $\vec{\varphi}_i$ of
each particle, $i$, of mass $m_i$ and tensorial moment of inertia, $\hat{J}_i$:
\begin{align}
  m_i\frac{\text{d}^2\vec{r}_i}{\text{d}t^2} &= \vec{F}_i = \sum\limits_{j\ne i} \vec{F}_{ij} + \vec{F}_i^\text{ext}\,,\label{eq:motion_mass_spring}\\
  \hat{J}_i\frac{\text{d}^2\vec{\varphi}_i}{\text{d}t^2} &=\vec{M}_i = \sum\limits_{j\ne i} \vec{M}_{ij}\,.
\end{align}
Here, $\vec{F}_i^\text{ext}$ is an external force, e.g. gravity, and $\vec{F}_{ij}$ and  $\vec{M}_{ij}$  are the force and torque acting on particle $i$ due to contacts with particles $j$. There are several models for $\vec{F}_{ij}$ and  $\vec{M}_{ij}$ as functions of the relative position, orientation, velocity and angular velocity of the involved particles, $i$ and $j$, for an extended discussion see, e.g., \cite{shaferForceSchemesSimulations1996,poschelComputationalGranularDynamics2005,kruggel-emdenReviewExtensionNormal2007,kruggel-emdenStudyTangentialForce2008,matuttisUnderstandingDiscreteElement2014}.


\subsection{Membrane: topology}
\label{sec: topology}

In the current paper, we focus on the description of an ambient membrane and its interaction with the granular particles. In our model, the elastically deformable membrane is modeled by mass-carrying particles that are connected by viscoelastic springs (mass-spring system, MSS). The topology of the membrane is given by a mathematical graph whose vertices and edges are represented by particles and springs, respectively.

\begin{figure}[htbp]
  \centering
	\includegraphics[width=0.95\linewidth]{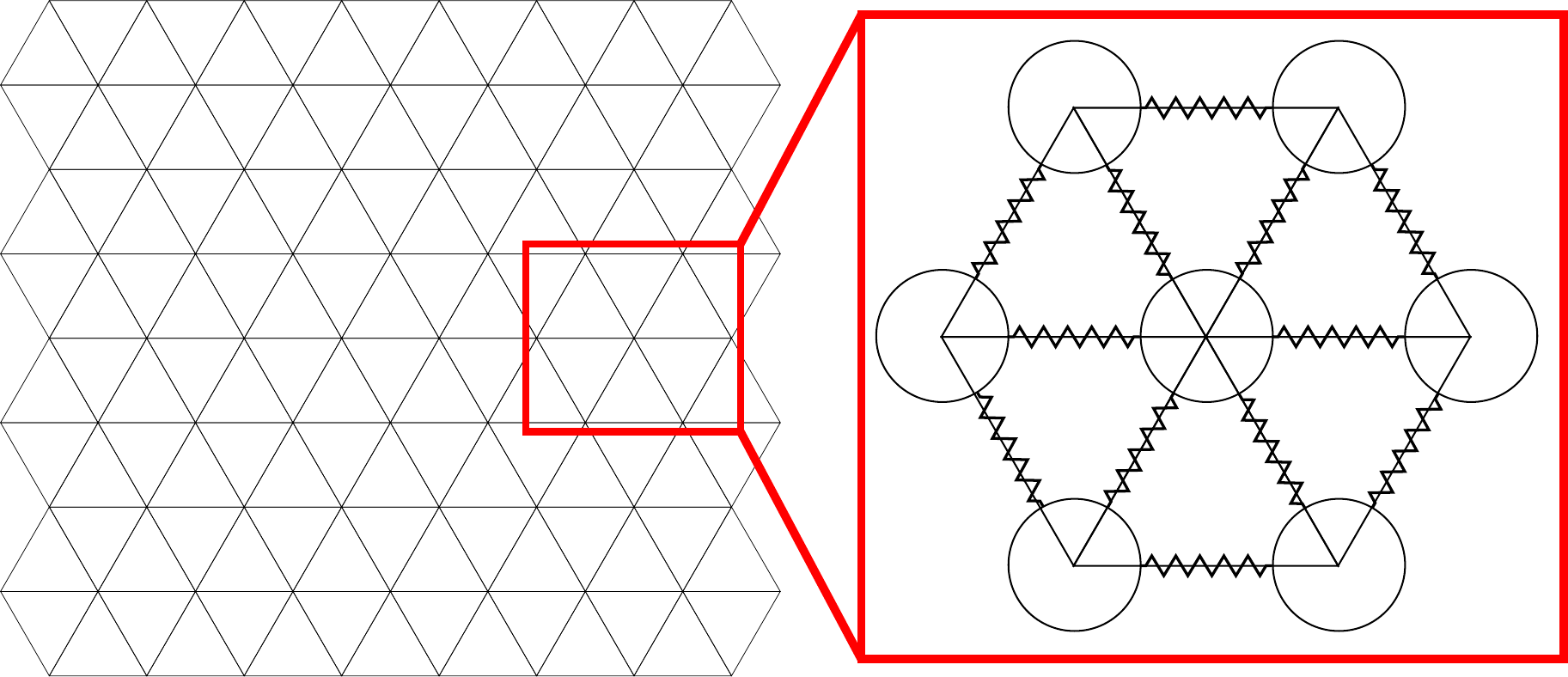}
	\caption{Sketch of the MSS  \label{img:setup_spring_bending}}
\end{figure}

The positions of the membrane particles (here called vertex particles), thus, describe the shape of the membrane (Fig. \ref{img:setup_spring_bending}).
They are subject of Newton's equations where the forces acting on the vertex membrane particles originate from three contributions: (a) viscoelastic stretching of the membrane, (b) moments due to bending of the membrane, and (c) interaction of granular particles with the membrane. We shall discuss these contributions in Secs. \ref{sec:stretch}-\ref{sec:membr-gran-membr}. The total force acting on the vertex particles is the sum of these three contributions.

\subsection{Membrane: stretching}
\label{sec:stretch}

Given two adjacent vertex particles $i,j$ at positions $\vec{\rho}_{i}$, $\vec{\rho}_{j}$ and velocities $\dot{\vec{\rho}}_{i}$, $\dot{\vec{\rho}}_{j}$, we define the relative quantities 
\begin{align}
  \vec{\rho}_{ij} &\equiv \vec{\rho}_{i} - \vec{\rho}_{j},\\
  \dot{\vec{\rho}}_{ij} &\equiv \dot{\vec{\rho}}_{i} - \dot{\vec{\rho}}_{j}
\end{align}
and the unit vector
\begin{align}
  \hat{\vec{\rho}}_{ij} &\equiv \frac{\vec{\rho}_{ij}}{\rho_{ij}}\,.
\end{align}
The interaction between particles $i$ and $j$ is due to linear elastic spring. Particle $i$ feels the force
\begin{align}
  \vec{F}_{ij}^{\,\text{spring}}
  = \hat{\vec{\rho}}_{ij}\left[ k\left(\rho_{ij} - \rho_{ij}^0\right) - 2\gamma
  \sqrt{k\,m_{ij}^\text{eff}} \,\hat{\vec{\rho}}_{ij}\cdot \dot{\vec{\rho}}_{ij}
  \right],
  \label{eq:mss_force}
\end{align}
which is the force of a damped harmonic oscillator, with equilibrium length $\rho_{ij}^0$, effective mass $m_{ij}^\text{eff}=m_im_j/(m_i+m_j)$, damping coefficient $\gamma$, and spring constant $k$.

To relate the spring constant, $k$, to material characteristics, we notice that each realistic membrane has a finite width, $d$, and the elasticity of the membrane material is characterized by its elastic modulus $E$. For our idealized two-dimensional membrane of vanishing thickness, one obtains the elastic constant \cite{kotMassSpringModels2017,ostoja-starzewskiLatticeModelsMicromechanics2002}.
\begin{equation}
  \label{eq:1}
k=\frac{\sqrt{3}}{2}E\,d\,.  
\end{equation}


\subsection{Membrane: flexibility}
\label{sec:membrane-pliability}

\begin{figure}[htbp]
  \centering
	\includegraphics[width=0.95\linewidth]{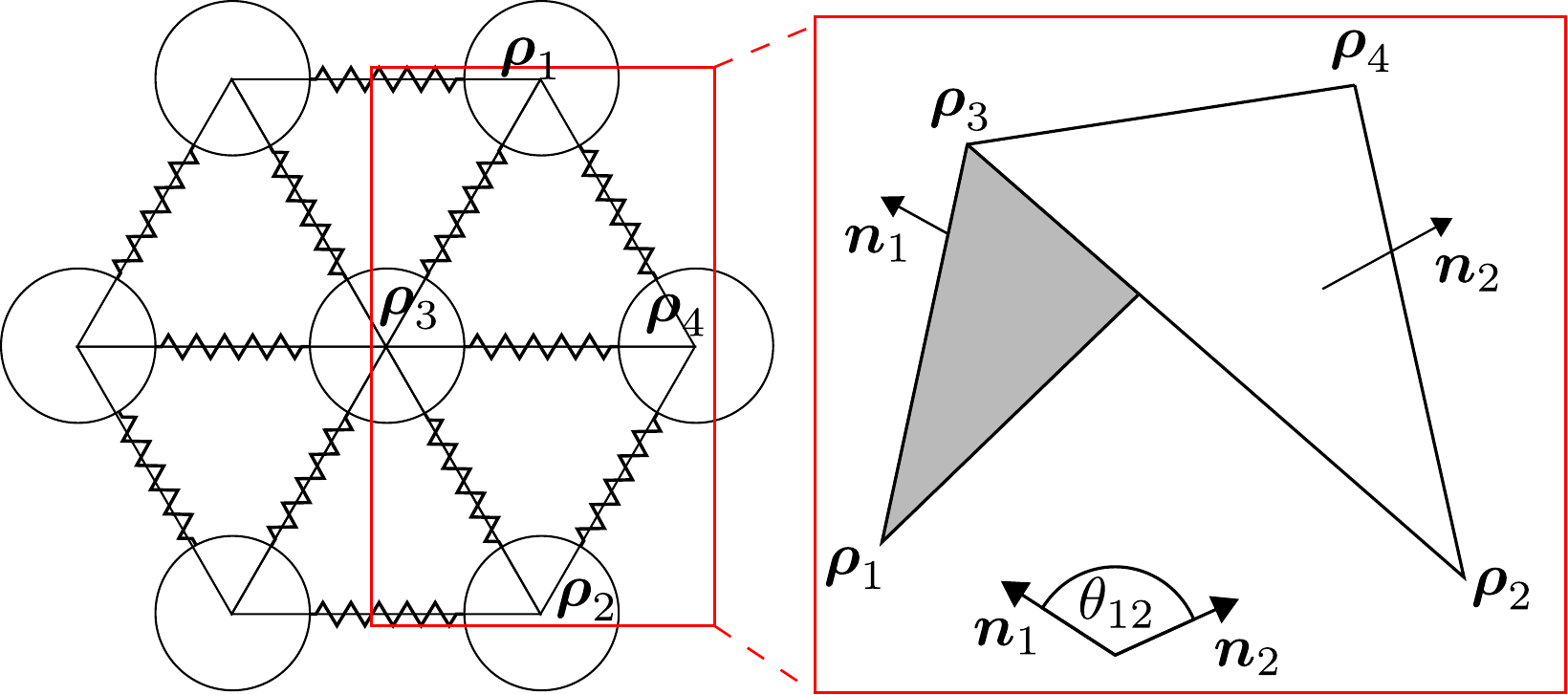}
	\caption{The deformation of the membrane is described by the angle $\theta_{12}$ between the normal vectors of adjacent triangles $\triangle134$ and $\triangle243$. $\theta_{12}$ is extremely exagerated in this sketch.
          \label{img:setup_bending}}
\end{figure}

To explain the description of membrane flexibility, we consider four adjacent vertex particles at positions $\vec{\rho}_{1}$, $\vec{\rho}_{2}$, $\vec{\rho}_{3}$, $\vec{\rho}_{4}$ \cite{bridsonSimulationClothingFolds2003}, see Fig. \ref{img:setup_bending}.
The vertex particles span two triangles with normal vectors
\begin{align}
  \label{eq:2}
  \vec{n}_1 &= (\vec{\rho}_1 - \vec{\rho}_3)\times (\vec{\rho}_1 - \vec{\rho}_4)\\
  \vec{n}_2&= (\vec{\rho}_2 - \vec{\rho}_4)\times (\vec{\rho}_2 - \vec{\rho}_3)\,.
\end{align}
The corresponding angle $\theta_{12}$
\begin{align}
  \label{eq:3}
  \cos \theta_{12}\equiv \vec{n}_1 \cdot \vec{n}_2
\end{align}
characterizes the flection of the triangles,
with respect to their common edge $\vec{\rho}_{43} = \vec{\rho}_4 - \vec{\rho}_3$. The restoring torque counteracting the flection can be expressed by elastic and dissipative forces, $\vec{F}_{i}^\text{el}$, acting on the involved vertex particles, $i\in\{1,2,3,4\}$
\cite{bridsonSimulationClothingFolds2003},
\begin{align}
    \vec{F}_{i}^\text{el} &= k^\text{el} \frac{\left|\vec{\rho}_{43}\right|^2}{|\vec{n}_1|+|\vec{n}_2|} \left(\sin{\frac{\theta_{12}}{2}}-\sin{\frac{\theta_{12}^0}{2}}\right)\vec{u}_i,\label{eq:bend:elastic}\\
  \vec{F}_{i}^\text{diss} &= -k^\text{diss}\,|\vec{\rho}_{43}|\,\dot{\theta}_{12}\vec{u}_i\,, \label{eq:bend:diss}
\end{align}
where $k^\text{el}$ and $k^\text{diss}$ are material parameters.

The directions of the forces are linear combinations of the triangles' normal vectors given by
\begin{align}
	\vec{u}_1 &= |\vec{\rho}_{43}|\frac{\vec{n}_1}{|\vec{n}_1|^2},\label{eq:bend:u1}\\
	\vec{u}_2 &= |\vec{\rho}_{43}|\frac{\vec{n}_2}{|\vec{n}_2|^2},\label{eq:bend:u2}\\
	\vec{u}_3 &= \frac{(\vec{x}_1-\vec{x}_4)\cdot\vec{\rho}_{43}}{|\vec{\rho}_{43}|}\frac{\vec{n}_1}{|\vec{n}_1|^2}
			   + \frac{(\vec{x}_2-\vec{x}_4)\cdot\vec{\rho}_{43}}{|\vec{\rho}_{43}|}\frac{\vec{n}_2}{|\vec{n}_2|^2},\label{eq:bend:u3}\\
   	\vec{u}_4 &= \frac{(\vec{x}_1-\vec{x}_3)\cdot\vec{\rho}_{43}}{|\vec{\rho}_{43}|}\frac{\vec{n}_1}{|\vec{n}_1|^2}
			   + \frac{(\vec{x}_2-\vec{x}_3)\cdot\vec{\rho}_{43}}{|\vec{\rho}_{43}|}\frac{\vec{n}_2}{|\vec{n}_2|^2}\,.\label{eq:bend:u4}
\end{align}
Each vertex particle is involved in 6 different pairs of triangles, see Fig. \ref{img:setup_spring_bending}. The total force acting on a vertex particle is, thus, the sum of the 6 corresponding forces given by Eqs. (\ref{eq:bend:elastic}, \ref{eq:bend:diss}).

\subsection{Membrane: granulate-membrane interaction}
\label{sec:membr-gran-membr}

\begin{figure}[htbp]
  \centering
	\includegraphics[width=0.95\linewidth]{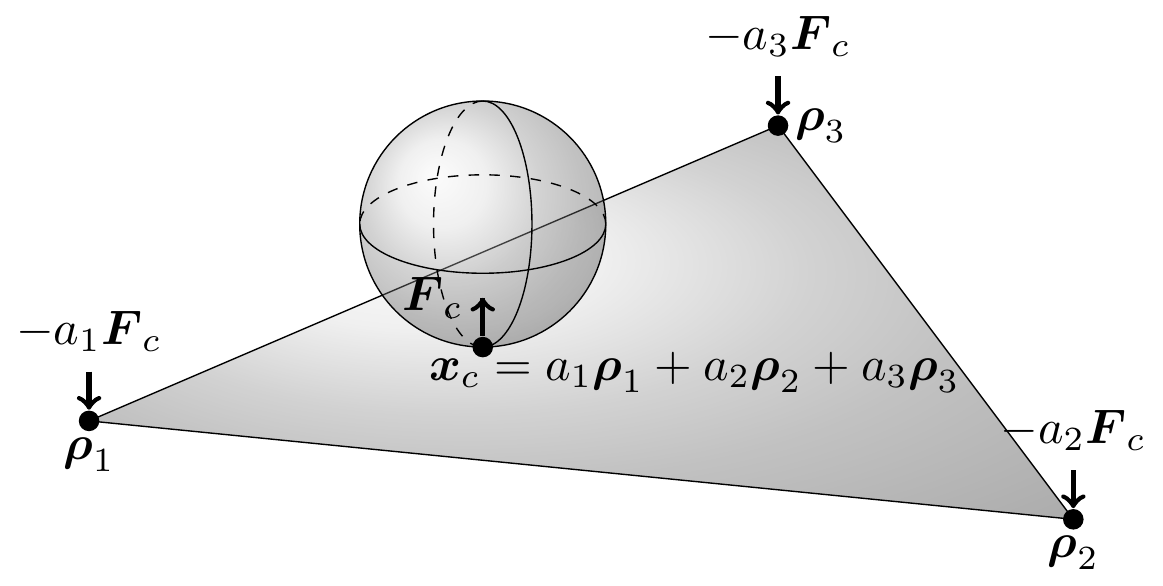}
	\caption{Sketch of a contact between a granular particle and a triangular patch. The resulting force $\vec{F}_c$ is mapped to the involved vertex particles $\vec{\rho}_{1}$, $\vec{\rho}_{2}$, $\vec{\rho}_{3}$ according to the barycentric weights $a_{1}$,  $a_{2}$, $a_{3}$ of the contact point $\vec{x}_c$ with respect to the locations of the vertex particles.
          \label{fig:model:barycentricInterpolation}}
\end{figure}

For the description of the interaction between the membrane and the confined granular particles, we assume triangular patches spanned between the time dependent momentary positions of adjacent vertex particles, Fig. \ref{fig:model:barycentricInterpolation}.
The interaction of the granular particles with the membrane is then described by contacts between the granular particles and the patches. This assures that the patches are always impenetrable disregarding of the sizes of the particles and the deformation of the membrane.

Contacts between a patch and a granular particle are classified as vertex, edge, or face contact:
\begin{equation}
\label{eq:5}
\begin{split}
\text{vertex contact if ~} &  \vec{x}_c=\vec{\rho}_i \,;~i\in\{1,2,3\}\\
\text{edge contact if~} &   \left(\vec{x}_c-\vec{\rho}_j\right)\times\left( \vec{\rho}_i-\vec{\rho}_j\right) = \vec{0} \,;\\
&   ~i,\,j\in\{1,2,3\} \,;~i \ne j \\
\text{face contact~} &  \text{else} 
  \end{split}
\end{equation}
In case a granular particle contacts two neighboring patches, $A$ and $B$, we chose the adequate contacts according to Tab. \ref{tab:choice} in dependence whether these contacts are face, edge, or vertex contacts \cite{huNewAlgorithmContact2013}
\begin{table}[h!]
  \centering
  \begin{tabular}{|c|c|c|c|}
    \hline
    & face $A$ & edge $A$ & vertex $A$ \\
    \hline
    face $B$ & $A$ and $B$ & $B$ & $B$ \\
    edge $B$ & $A$ & $A$ or $B$ (rand) & $B$\\
    vertex $B$ & $A$ & $A$ & $A$ or $B$ (rand)\\
    \hline
  \end{tabular}
  \caption{Selection of contacts for the case that a granular particle ist in contact with two adjacent patches, $A$ and $B$}
  \label{tab:choice}
\end{table}
These selection rules do only apply if the patches $A$ and $B$ have a common edge. Otherwise, all contacts are handles regularly.

%

In case a granular particle contacts three neighboring patches at their common vertex, one of these contacts is selected randomly and the others are disregarded.

Contacts of the membrane with itself may be calculated from contacts between vertex particles and triangular patches.

  Once the contact point is defined, we compute the force according to the specified contact law. The relative velocity of the granular particle and the membrane at the contact point which enters the force is interpolated from the velocities of the vertex particles using barycentric weights, as sketched in Fig. \ref{fig:model:barycentricInterpolation}. Similarly, the obtained force is distributed to the involved vertex particles with barycentric weights $a_1$, $a_2$, $a_3$. The positions and velocities of the vertex particles and, thus, the dynamics of the membrane are obtained by numerical integration in the same way as the granular particles.

The above selection rule above in combination with the barycentric partition of the force leads to smooth and physically plausible forces acting on the vertex particles. \cite{huNewAlgorithmContact2013}

\section{Applications}

We implemented the described flexible wall into the DEM program MercuryDPM \cite{weinhartFastFlexibleParticle2020}. Here we present two examples of its application.

\subsection{Triaxial test}\label{sec:res:triaxialTest}
\begin{figure}[htbp]
    \includegraphics[width=0.49\linewidth]{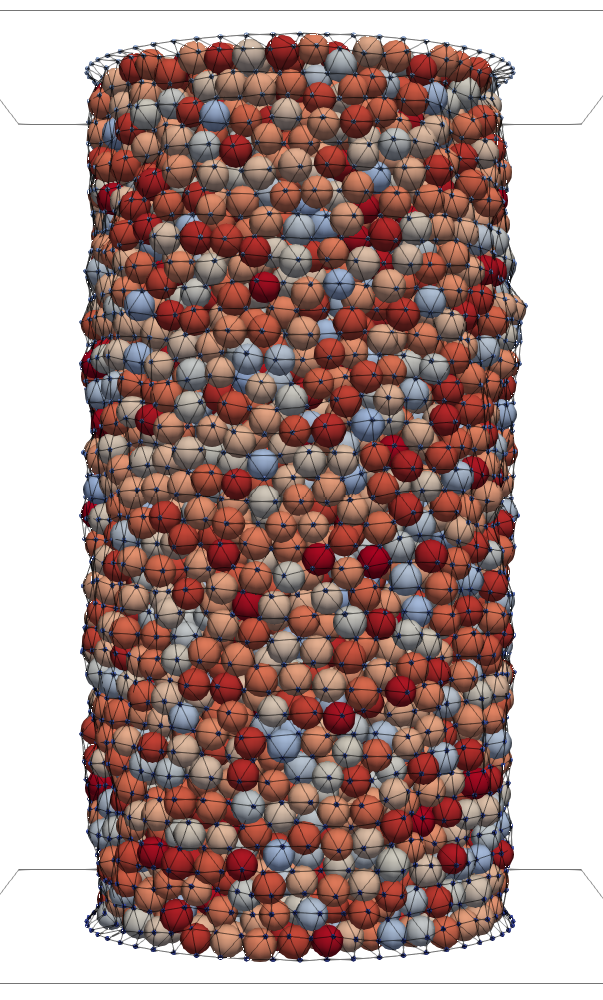}
    \includegraphics[width=0.49\linewidth]{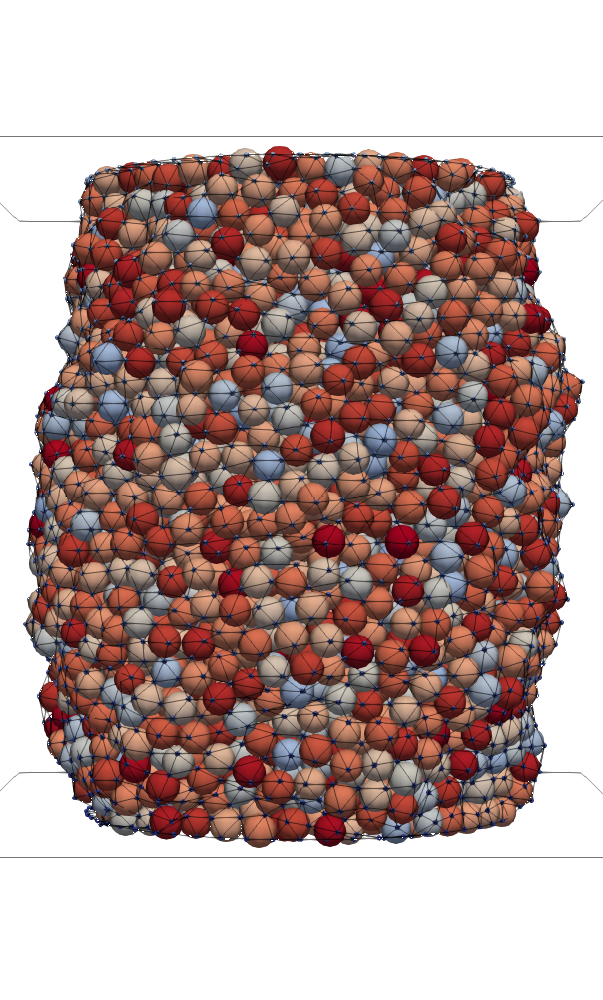}
    \caption{A triaxial test cell in its initial and final states}
    \label{fig:triaxialTest:image}
\end{figure}

The triaxial test is commonly used to investigate the mechanical properties of a granular sample. To that aim, the sample is placed between two parallel platens and wrapped by a cylindrical membrane, see Fig. \ref{fig:triaxialTest:image}.
A confining stress $\sigma_\text{c}$ in radial direction is applied through the membrane. By controlling the initial distance $h=h_0$ between the platens, an initial stress $\sigma_\text{p}=\sigma_\text{c}$ is applied in axial direction. Then, the platens are displaced at constant relative velocity $\vec{v_p}$ to apply a strain $\epsilon=\log(h_0/h)$. We record the corresponding deviatoric stress $\sigma_\text{d} = \sigma_\text{p}-\sigma_\text{c}$.

In the simulation, we represent the platens by rigid walls and the membrane by an MSS. We place small particles at random positions inside the membrane such that they do not contact one another. To generate the initial conditions, we apply the Lubachevsky–Stillinger algorithm \cite{lubachevskyGeometricPropertiesRandom1990}. We then gradually apply an initial stress, $\sigma_p$, to the platens and a confining stress, $\sigma_\text{c}$, to the membrane. To this end, each triangular patch of area $A^w_i$ and normal unit vector $\hat{\vec{n}}^{\,w}_i$ is loaded with a force
\begin{equation}
  \label{eq:4}
  \vec{F}^{\,w}_i=A^{\,w}_i\sigma_\text{c}\,\hat{\vec{n}}^{\,w}_i  \,,
\end{equation}
where $\hat{\vec{n}}^{\,w}_i$ is defined such that force acts from outside the membrane to the granulate located inside. After this initialization, we displace the platens at relative velocity $\vec{v}_\text{p}$ and record the deviatoric stress, $\sigma_\text{d}$. Figure \ref{fig:triaxialTest:image} shows the initial and final states of such a simulation.

To demonstrate the performance of our model, we perform simulations for two different cases:
\begin{enumerate}
\item We describe the membrane by a MSS where the confining stress of the membrane is provided by contacts between vertex particles and  granular particles. This approach was used before, e.g., in \cite{quDiscreteElementModelling2019}. We term this setup \emph{$\text{MSS}_\text{particle}$}. 
\item We describe the membrane as described in Sec. \ref{sec:model-description}. Here the confining stress of the membrane is provided by contacts  between the granular particles and the triangular patches. We term this setup \emph{$\text{MSS}_\text{patch}$}. 
\end{enumerate}
For $\text{MSS}_\text{particle}$, we place the vertex particles of the membrane spaced by $s=1.33\,\text{mm}$, such that the radius of a vertex particle is about $1/3$ the radius of a granular particle. This value is a trade-off between keeping the computational cost low and having a smooth membrane that prevents penetration of granular particles \cite{debonoDiscreteElementModelling2012,quDiscreteElementModelling2019}. For $\text{MSS}_\text{patch}$, we use $s=4.0\,\text{mm}$ because the membrane has a smooth surface and is impenetrable by design.

We perform the triaxial test at velocity $v_\text{p}=0.05\,\text{m}/\text{s}$ and confining pressure $\sigma_\text{c}=100\,\text{kPa}$. The material parameters are given in Tab. \ref{tab:matProp}. Furthermore, we choose the parameters $\gamma\approx0.15$, $k^\text{el}=10^{-3}\,\text{N/m}$ and $k^\text{diss}=10^{-4}\,\text{Ns/m}$. Using the initial areas of the triangles, we use Eqs. (\ref{eq:bend:elastic}-\ref{eq:bend:u4}) to compute the confining forces.

\begin{table}[htbp] 
\caption{Material parameters used in the simulation\label{tab:matProp}\label{tab1}}
\begin{tabularx}{\columnwidth}{lll}
\toprule
 & particles	& membrane\\
\midrule
     radius / thickness [mm] &  $2.3$ to $2.7$ & 0.3 \\
     elastic modulus [Pa] & $4.6\cdot10^{10}$ & $1.25\cdot10^6$\\
     Poisson's ratio & 0.245 & 1/3\\
     friction coefficient & 0.25 & 1.2\\
\bottomrule
\end{tabularx}
\end{table}

We consider two different systems: \emph{case 1} -- a cylinder of initial height $100\,\text{mm}$ and radius $25\,\text{mm}$, and \emph{case 2} - a cylinder of initial height $140\,\text{mm}$ and radius $35\,\text{mm}$. The numbers of particles used for the membrane and the granulate are given in Tab. \ref{tab:triaxialTest:numberOfParticles}.
\begin{table}[htbp] 
  \caption{Number of granular and membrane particles in the simulations}
  \begin{tabularx}{\columnwidth}{lllll}
  \toprule
   & \multicolumn{2}{c}{membrane particles} & \multicolumn{2}{c}{granular particles} \\
   & case 1 & case 2 & case 1 & case 2\\
  \midrule
       $\text{MSS}_\text{particle}$ & 10440 & 20374 & $1690$ & $4790$ \\
       $\text{MSS}_\text{patch}$ & 1131 & 2214 & $1720$ & $4670$\\     
  \bottomrule
  \end{tabularx}
  \label{tab:triaxialTest:numberOfParticles}
\end{table}

\begin{figure}[tbp]
  \subfigure[Case 1.  initial height $100\,\text{mm}$; radius $25\,\text{mm}$
  \label{fig:triaxialTest:stressStrainCase1}]{\includegraphics[width=0.99\linewidth]{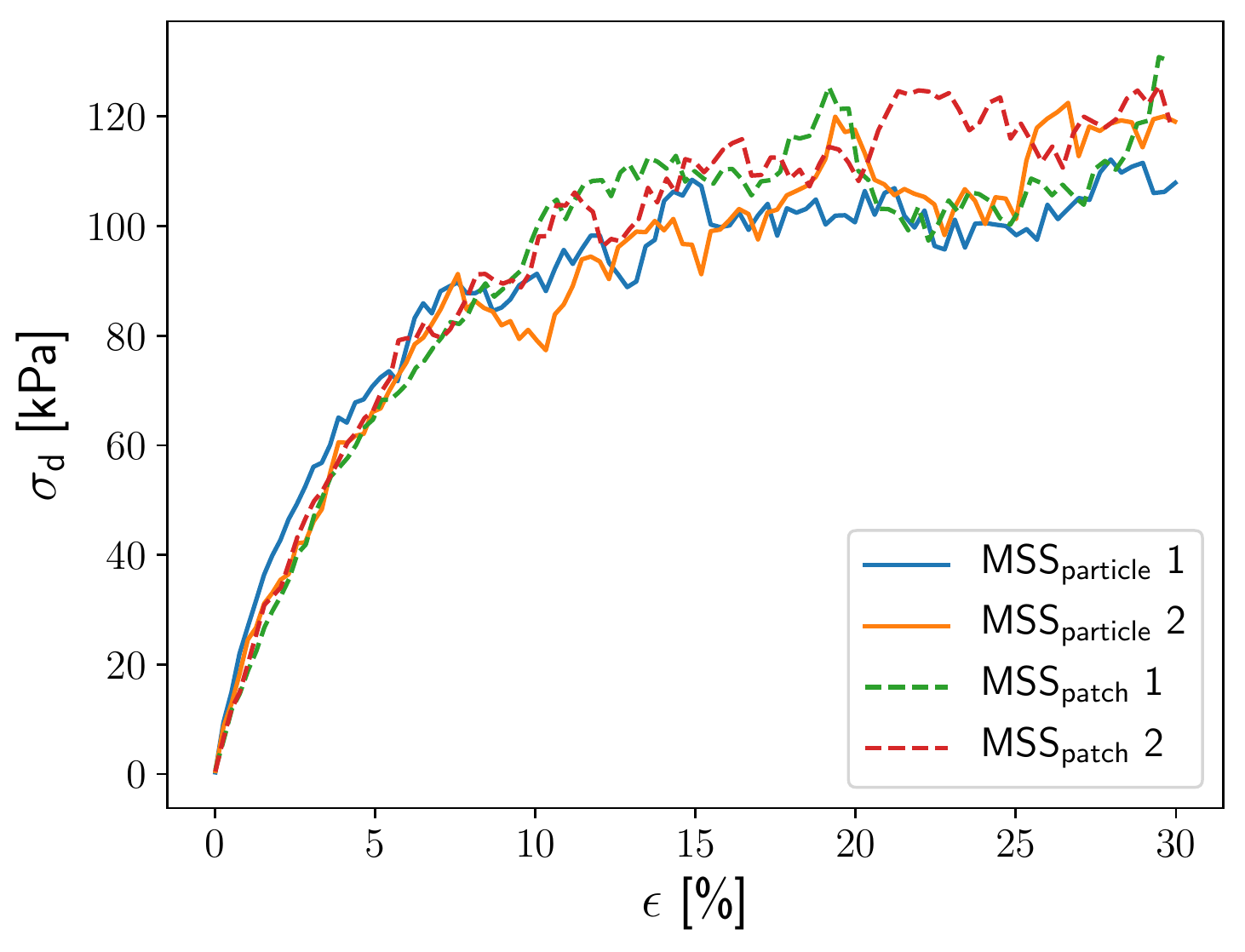}}
    \subfigure[Case 2. initial height $140\,\text{mm}$; radius $35\,\text{mm}$
    \label{fig:triaxialTest:stressStrainCase2}]{\includegraphics[width=0.99\linewidth]{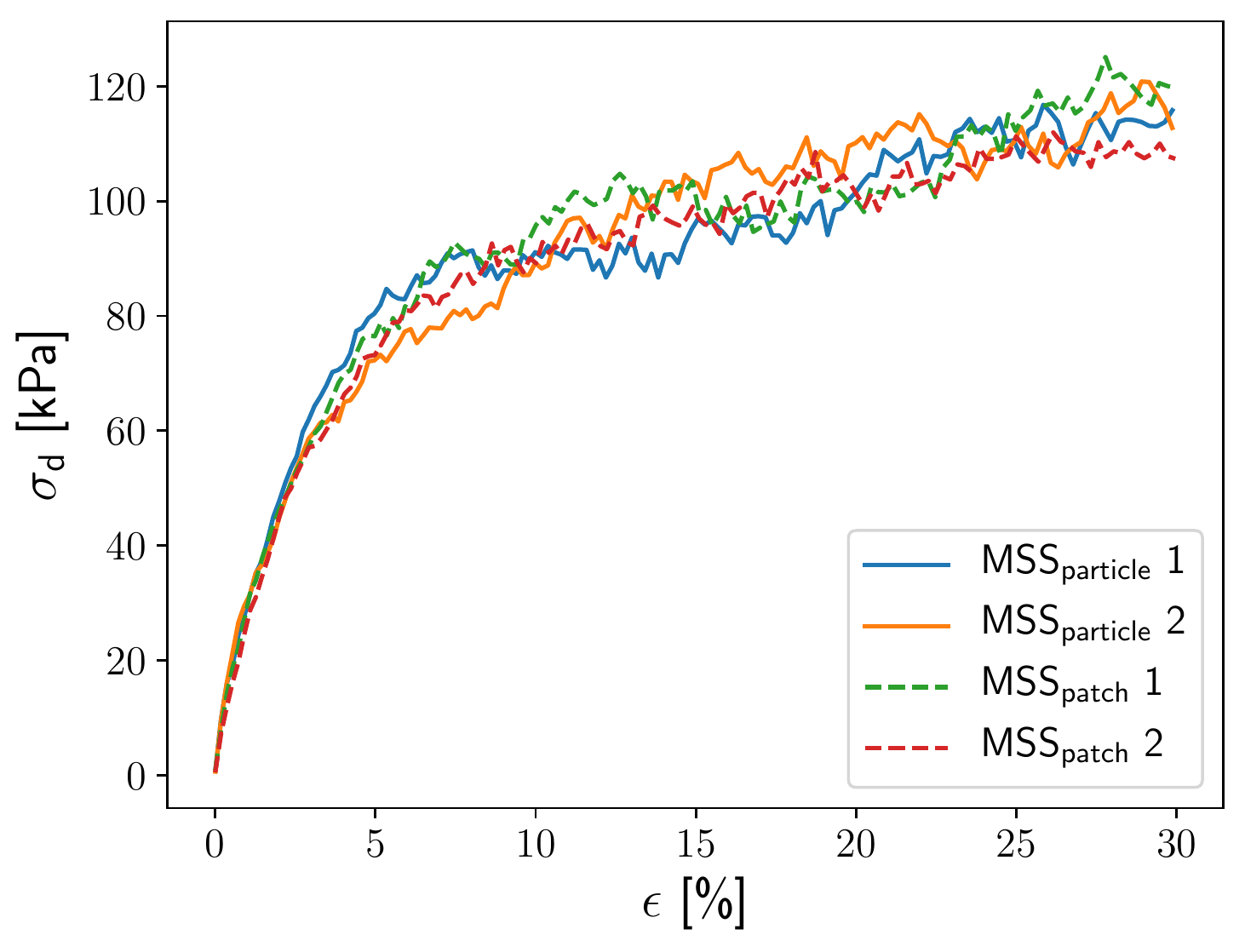}}
    \caption{Deviatoric stress against axial strain for two different geometries}
    \label{fig:triaxialTest:stressStrain}
\end{figure}

Figure \ref{fig:triaxialTest:stressStrain}
shows the deviatoric stress, $\sigma_\text{d}$, as a function of the axial strain, $\epsilon$. The different membrane representations, $\text{MSS}_\text{particle}$ and $\text{MSS}_\text{patch}$, do not lead to significant differences of the stress-strain behavior.

Table \ref{tab:triaxialTest:exec_time} compares the computer time used to simulate a real time of $20\,\text{ms}$.
\begin{table}[htbp] 
\centering
\caption{Computer time used for different membrane representations. Real time is $20\,\text{ms}$}
\begin{tabularx}{\columnwidth}{lll}
\toprule
 & case 1	& case 2\\
\midrule
     $\text{MSS}_\text{particle}$ & $\approx32\,\text{min}$ & $\approx96\,\text{min}$ \\
     $\text{MSS}_\text{patch}$ & $\approx5\,\text{min}$ & $\approx17\,\text{min}$\\     
\bottomrule
\end{tabularx}
\label{tab:triaxialTest:exec_time}
\end{table}
The reduced number of particles in $\text{MSS}_\text{patch}$  accelerates the simulations by about the factor 5, compared to simulations using $\text{MSS}_\text{particle}$.

\subsection{Friction test}
A correct representation of frictional forces at contacts with the membrane is important for many applications. For instance, one of the mechanisms allowing a granular gripper to grasp an object relies on frictional forces \cite{brownUniversalRoboticGripper2010}. By construction, an object gliding on a membrane modeled by particles cannot feel a constant (or, at least, smnooth) force.  

We demonstrate the smoothness of a membrane modeled by the here described model, and the resulting frictional forces by means of a simple sliding test: In $\text{MSS}_\text{particle}$, a membrane is modeled by vertex particles with spacing $1.33\,\text{mm}$. In $\text{MSS}_\text{patch}$, the membrane is modeled by patches of side length $1.33\,\text{mm}$.

\begin{figure}[htbp]
    \centering
    \includegraphics[width=0.95\linewidth]{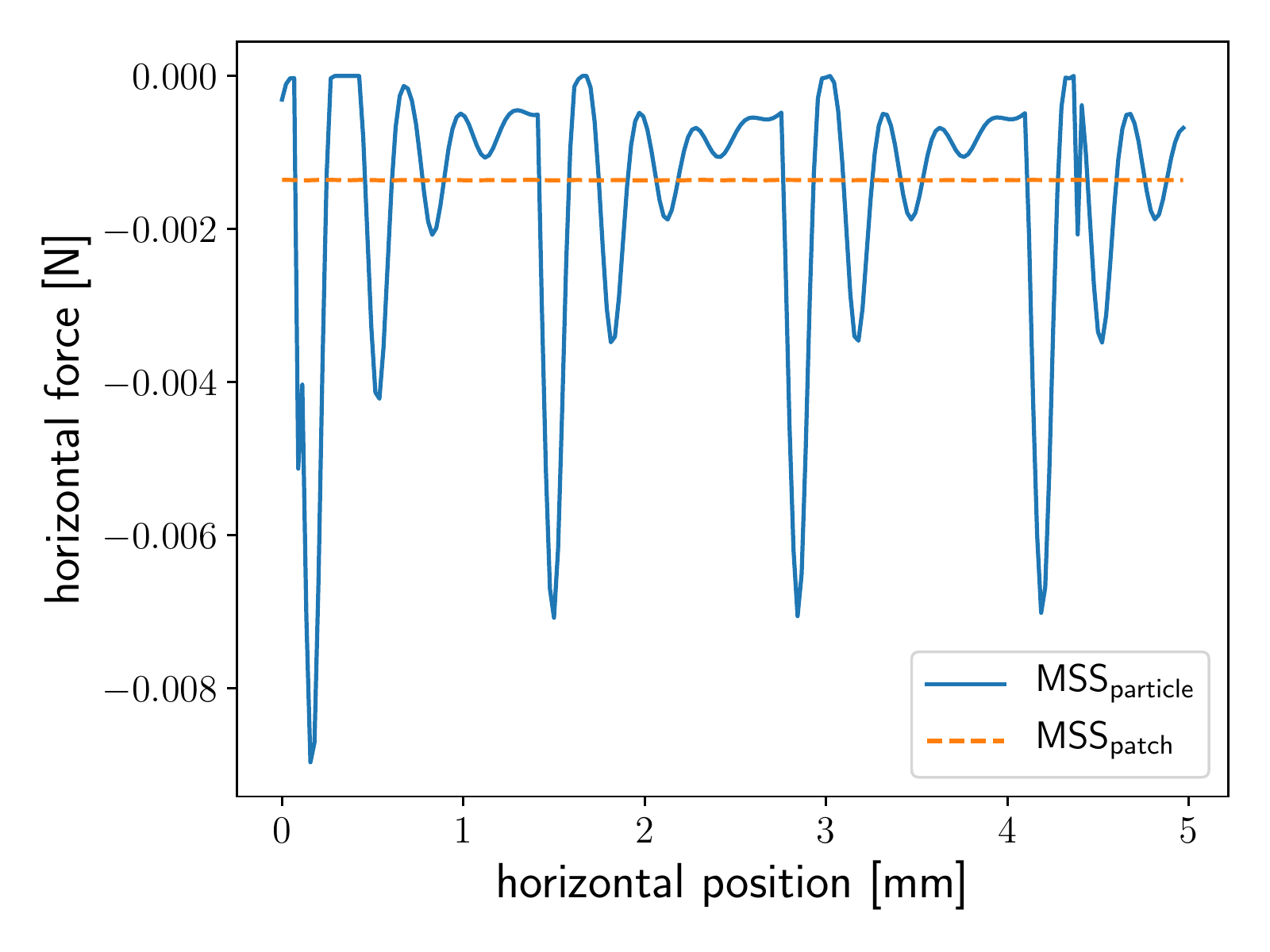}
    \includegraphics[width=0.95\linewidth]{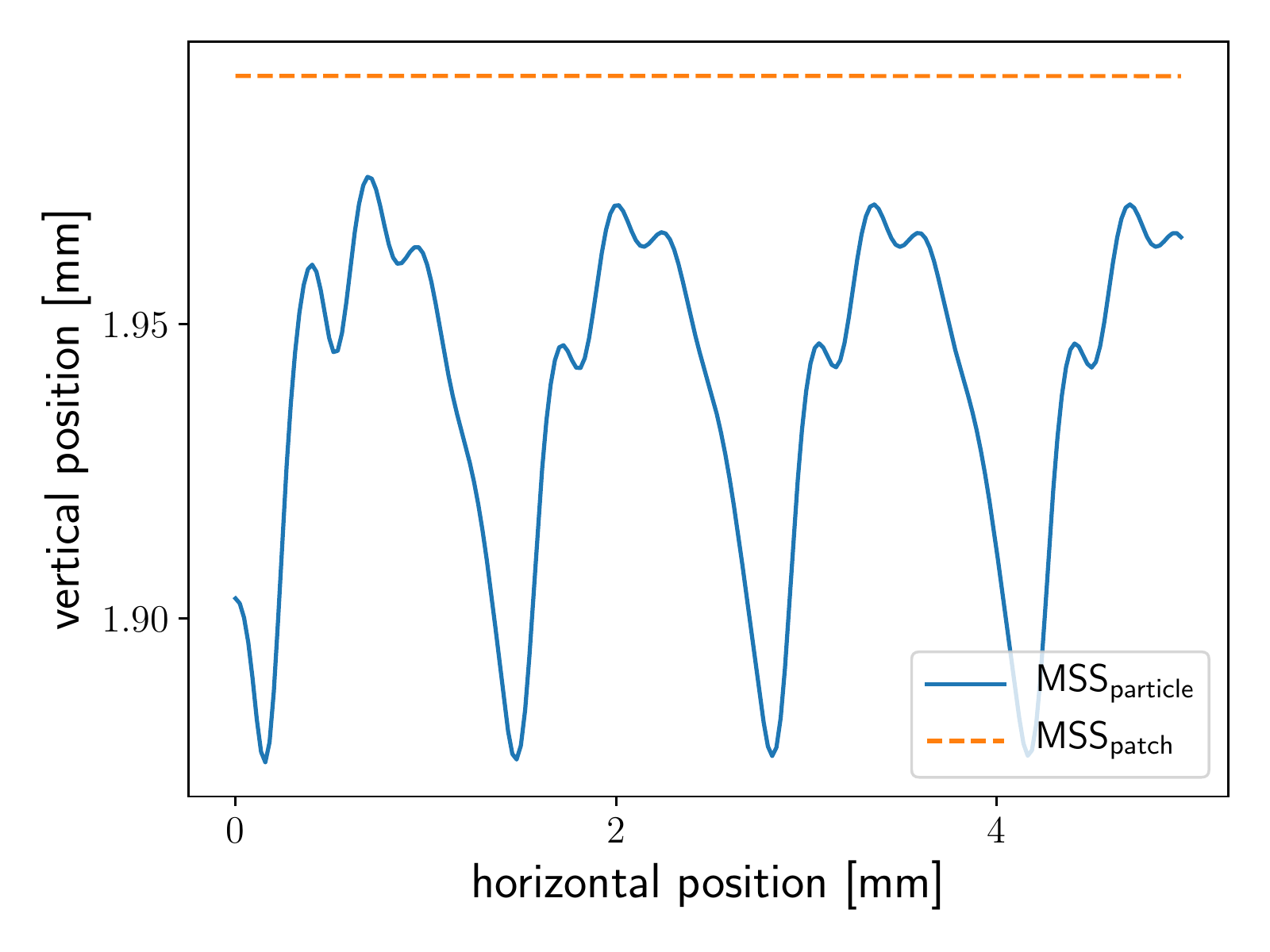}
    \caption{The frictional force in horizontal direction felt a particle when moving on the surface of a membrane (upper image) and corresponding vertical position (lower image) vs the horizontal position for both considered membrane models. Rolling degree of freedom was suppressed}
    \label{fig:slidingTest:forcePosition}
\end{figure}

For the test, we place a spherical particle of radius $2\,\text{mm}$ on a membrane, the free motion of this particle is restricted to the vertical coordinate, perpendicular to the membrane. Its horizontal motion at constant velocity, $0.05\,\text{m/s}$,  is enforced externally. Figure \ref{fig:slidingTest:forcePosition}  shows the particle's vertical position and the frictional force in sliding direction.

For $\text{MSS}_\text{particle}$ we see oscillations in both horizontal force and vertical position. For $\text{MSS}_\text{patch}$, we instead observe constant vertical position and constant friction force, as expected when a particle slides on an even plane.

\section{Conclusions}
Previous work using MSS for representing membranes within DEM simulations describe contacts between the granulate and confining membranes through contacts between the membrane's particles and the particles of the granulate. In this paper, we introduced a membrane model using surface patches. Contacts between granuate and the membrane are described by contacts between surface patches constituting the membrane and the particles of the granulate.

The movel model describes a closed surface by design, therefore, the number of particles in the MSS can be reduced drastically. In a sample simulation modeling a triaxial test, we obtained an acceleration of the mumerical method by about a factor five. The comparison of our results with the results using a traditional membrane description did not reveal significant differences in the physical behavior. We further demonstrated the new model's ability to represent smooth surfaces.

\begin{acknowledgements}
We gratefully acknowledge funding by the Deutsche Forschungsgemeinschaft (DFG, German Research Foundation) – project number 411517575. 
The work was supported by the Interdisciplinary Center for Nanostructured Films (IZNF), the Competence Unit for Scientific Computing (CSC), and the Interdisciplinary Center for Functional Particle Systems (FPS) at Friedrich-Alexander Universit\"at Erlangen-N\"urnberg. 
  
\end{acknowledgements}

\section*{Conflict of interest}
The authors declare that they have no conflict of interest.

\bibliographystyle{spmpsci}      
\bibliography{membrane-bib.bib}   

\begin{thebibliography}{10}
\providecommand{\url}[1]{{#1}}
\providecommand{\urlprefix}{URL }
\expandafter\ifx\csname urlstyle\endcsname\relax
  \providecommand{\doi}[1]{DOI~\discretionary{}{}{}#1}\else
  \providecommand{\doi}{DOI~\discretionary{}{}{}\begingroup
  \urlstyle{rm}\Url}\fi

\bibitem{liuJammingNotJust1998}
Liu, A. J., Nagel, S. R.: Jamming is not just cool any more.
\newblock Nature \textbf{396}, 21--22 (1998).
\newblock \doi{10.1038/23819}

\bibitem{ohernJammingZeroTemperature2003}
O’Hern, C. S., Silbert, L. E., Liu, A. J., Nagel, S. R.: Jamming at zero
  temperature and zero applied stress: {{The}} epitome of disorder.
\newblock Phys. Rev. E \textbf{68}, 011306 (2003).
\newblock \doi{10.1103/PhysRevE.68.011306}

\bibitem{ciamarraRecentResultsJamming2010}
Ciamarra, M. P., Nicodemi, M., Coniglio, A.: Recent results on the jamming
  phase diagram.
\newblock Soft Matter \textbf{6}, 2871--2874 (2010).
\newblock \doi{10.1039/B926810C}

\bibitem{brownUniversalRoboticGripper2010}
Brown, E., Rodenberg, N., Amend, J., Mozeika, A., Steltz, E., Zakin, M. R.,
  Lipson, H., Jaeger, H. M.: Universal robotic gripper based on the jamming of
  granular material.
\newblock PNAS \textbf{107}, 18809--18814 (2010).
\newblock \doi{10.1073/pnas.1003250107}

\bibitem{hauserFrictionDampingCompliant2016}
Hauser, S., Eckert, P., Tuleu, A., Ijspeert, A.: Friction and damping of a
  compliant foot based on granular jamming for legged robots.
\newblock In: 2016 6th {{IEEE International Conference}} on {{Biomedical
  Robotics}} and {{Biomechatronics}} ({{BioRob}}), pp. 1160--1165 (2016).
\newblock \doi{10.1109/BIOROB.2016.7523788}

\bibitem{fitzgeraldReviewJammingActuation2020}
Fitzgerald, S. G., Delaney, G. W., Howard, D.: A {{Review}} of {{jamming
  actuation}} in {{soft robotics}}.
\newblock Actuators \textbf{9}, 104 (2020).
\newblock \doi{10.3390/act9040104}

\bibitem{debonoDiscreteElementModelling2012}
{de Bono}, J., Mcdowell, G., Wanatowski, D.: Discrete element modelling of a
  flexible membrane for triaxial testing of granular material at high
  pressures.
\newblock Geotech. Lett. \textbf{2}, 199--203 (2012).
\newblock \doi{10.1680/geolett.12.00040}

\bibitem{quDiscreteElementModelling2019}
Qu, T., Feng, Y. T., Wang, Y., Wang, M.: Discrete element modelling of flexible
  membrane boundaries for triaxial tests.
\newblock Comput. Geotech. \textbf{115}, 103154 (2019).
\newblock \doi{10.1016/j.compgeo.2019.103154}

\bibitem{wuStudyShearBehavior2021}
Wu, K., Sun, W., Liu, S., Zhang, X.: Study of shear behavior of granular
  materials by {{3D DEM}} simulation of the triaxial test in the membrane
  boundary condition.
\newblock Adv. Powder Technol. \textbf{32}, 1145--1156 (2021).
\newblock \doi{10.1016/j.apt.2021.02.018}

\bibitem{gotzSoftParticlesReinforce2021}
G{\"o}tz, H., Santarossa, A., Sack, A., P{\"o}schel, T., M{\"u}ller, P.: Soft
  particles reinforce robotic grippers: Robotic grippers based on granular
  jamming of soft particles.
\newblock Granular Matter \textbf{24}, 31 (2021).
\newblock \doi{10.1007/s10035-021-01193-4}

\bibitem{bendingBeamMeta:2022}
G{\"o}tz, H., P{\"o}schel, T.: Granular meta-material: {V}iscoelastic response
  of a bending beam.
\newblock Granular Matter \textbf{sumitted} (2022)

\bibitem{shaferForceSchemesSimulations1996}
Sch{\"a}fer, J., Dippel, S., Wolf, D. E.: Force {{schemes}} in {{simulations}}
  of {{granular materials}}.
\newblock J. Phys. I France \textbf{6}, 5--20 (1996).
\newblock \doi{10.1051/jp1:1996129}

\bibitem{poschelComputationalGranularDynamics2005}
P{\"o}schel, T., Schwager, T.: Computational Granular Dynamics: Models and
  Algorithms.
\newblock {Springer-Verlag}, {Berlin ; New York} (2005)

\bibitem{kruggel-emdenReviewExtensionNormal2007}
{Kruggel-Emden}, H., Simsek, E., Rickelt, S., Wirtz, S., Scherer, V.: Review
  and extension of normal force models for the {{Discrete Element Method}}.
\newblock Powder Technology \textbf{171}, 157--173 (2007).
\newblock \doi{10.1016/j.powtec.2006.10.004}

\bibitem{kruggel-emdenStudyTangentialForce2008}
{Kruggel-Emden}, H., Wirtz, S., Scherer, V.: A study on tangential force laws
  applicable to the discrete element method ({{DEM}}) for materials with
  viscoelastic or plastic behavior.
\newblock Chemical Engineering Science \textbf{63}(6), 1523--1541 (2008).
\newblock \doi{10.1016/j.ces.2007.11.025}

\bibitem{matuttisUnderstandingDiscreteElement2014}
Matuttis, H. G., Chen, J.: Understanding the {{Discrete Element Method}}:
  {{Simulation}} of {{Non-Spherical Particles}} for {{Granular}} and
  {{Multi-body Systems}}.
\newblock {Wiley} (2014)

\bibitem{kotMassSpringModels2017}
Kot, M., Nagahashi, H.: Mass spring models with adjustable {{Poisson}}'s ratio.
\newblock Vis. Comput. \textbf{33}, 283--291 (2017).
\newblock \doi{10.1007/s00371-015-1194-8}

\bibitem{ostoja-starzewskiLatticeModelsMicromechanics2002}
{Ostoja-Starzewski}, M.: Lattice models in micromechanics.
\newblock Appl. Mech. Rev. \textbf{55}, 35--60 (2002).
\newblock \doi{10.1115/1.1432990}

\bibitem{bridsonSimulationClothingFolds2003}
Bridson, R., Marino, S., Fedkiw, R.: Simulation of clothing with folds and
  wrinkles.
\newblock In: Proceedings of the 2003 {{ACM SIGGRAPH}}/{{Eurographics}}
  Symposium on {{Computer}} Animation, {{SCA}} '03, pp. 28--36. {Eurographics
  Association}, {Goslar, DEU} (2003)

\bibitem{huNewAlgorithmContact2013}
Hu, L., Hu, G. M., Fang, Z. Q., Zhang, Y.: A new algorithm for contact
  detection between spherical particle and triangulated mesh boundary in
  discrete element method simulations.
\newblock Int. J. Numer. Methods Eng. \textbf{94}, 787--804 (2013).
\newblock \doi{10.1002/nme.4487}

\bibitem{weinhartFastFlexibleParticle2020}
Weinhart, T., Orefice, L., Post, M., {van Schrojenstein Lantman}, M. P.,
  Denissen, I. F. C., Tunuguntla, D. R., Tsang, J. M. F., Cheng, H., Shaheen,
  M. Y., Shi, H., Rapino, P., Grannonio, E., Losacco, N., Barbosa, J., Jing,
  L., Alvarez~Naranjo, J. E., Roy, S., {den Otter}, W. K., Thornton, A. R.:
  Fast, flexible particle simulations \textemdash{} {{An}} introduction to
  {{MercuryDPM}}.
\newblock Comput. Phys. Commun. \textbf{249}, 107129 (2020).
\newblock \doi{10.1016/j.cpc.2019.107129}

\bibitem{lubachevskyGeometricPropertiesRandom1990}
Lubachevsky, B. D., Stillinger, F. H.: Geometric properties of random disk
  packings.
\newblock J. Stat. Phys. \textbf{60}, 561--583 (1990).
\newblock \doi{10.1007/BF01025983}

\end{thebibliography}

\end{document}